\documentclass{easychair}

\usepackage{doc}
\usepackage{xspace}
\usepackage{tikz}
\usepackage[T1]{fontenc}
\usetikzlibrary{shapes,arrows}
\usepackage{smartdiagram}
\usesmartdiagramlibrary{additions}
\usepackage{array}
\usepackage{subcaption}

\newcommand{\symbiotic}{\textsc{Symbiotic}\xspace}
\newcommand{\clang}{\textsc{clang}\xspace}
\newcommand{\instr}{\texttt{sbt-instrumentation}\xspace}

\title{ SBT-instrumentation: A Tool for Configurable\\Instrumentation of LLVM Bitcode }

\author{ Martina Vitovsk\'{a} \and
Marek Chalupa \and Jan Strej\v{c}ek }

\institute{
  Masaryk University,
  Brno, Czech Republic\\
  \email{\{xvitovs1,\,xchalup4,\,strejcek\}@fi.muni.cz}
 }

\authorrunning{Vitovsk\'{a}, Chalupa, Strej\v{c}ek}

\titlerunning{SBT-instrumentation}

\begin{document}

\maketitle

\begin{abstract}
  The paper describes a member of the \symbiotic toolbox called
  \instr, which is a tool for configurable instrumentation of LLVM
  bitcode. The tool enables a user to specify patterns of instructions
  and to define functions whose calls will be inserted before or after
  instructions that match the patterns. Moreover, the tool offers
  additional functionality. First, the instrumentation can be divided
  into phases in order to pass information acquired in an earlier
  phase to the later phases. Second, it can utilize results of some
  external static analysis by connecting it as a plugin. The \instr
  tool has been developed as the part of \symbiotic responsible for
  inserting memory safety checks. However, its configurability
  opens the way to use it for many various purposes.
\end{abstract}

% The table of contents below is added for your convenience. Please do not use
% the table of contents if you are preparing your paper for publication in the
% EPiC Series or Kalpa Publications series

%\setcounter{tocdepth}{2}
%{\small
%\tableofcontents}

%------------------------------------------------------------------------------
\section{Introduction}
\label{sect:introduction}

Instrumentation is a technique widely used in the field of program analysis and
verification that inserts an auxiliary code into the program that is being
analyzed. A number of well-known tools use instrumentation to achieve their
goals. For example, \clang's AddressSanitizer~\cite{Serebryany2012} uses
compile-time instrumentation to check for memory errors,
JProfiler~\cite{JProfiler} uses instrumentation for profiling of Java
applications, and CBMC~\cite{cbmc} inserts code checking various defects like
overflows or memory errors. However, the tools use mostly a single-purpose
instrumentation that cannot be easily modified. In this paper, we present a
tool \instr~\cite{thesis} that offers general instrumentation of LLVM
bitcode~\cite{llvm} that is easily configurable according to the user's choice. 

The tool was primarily designed for inserting memory safety checks
that can detect invalid dereferences, invalid deallocations, and
memory leaks. Roughly speaking, memory safety checking needs to insert
two types of functions:
\begin{itemize}
\item tracking functions that build and maintain records about
  allocated memory blocks and
\item checking functions that use these records to decide the validity
  of memory operations.
\end{itemize}
This initial application~\cite{spin2018} brought several interesting and non-standard
features to our tool. First, insertion of many memory safety checks
can be avoided as a static pointer analysis can guarantee that many
dereferences and deallocations are safe. Hence, \instr supports
plugins performing pointer or other analyses and insertions of checks
can be conditioned by an output of these plugins. Further, one needs
to track only memory blocks relevant for inserted checks. Thus, the
instrumentation first inserts checks and only then the necessary
tracking functions are inserted. This motivated performing
instrumentation in phases, where subsequent phases can use information
gathered in the previous phases.
The ability to use static analysis and phased instrumentation has led
to a dramatic reduction of the number of inserted instructions
by more than $85\%$ in our experiments~\cite{spin2018}.
This reduction can make the instrumented code more efficient when executed
or easier for subsequent analysis or verification.

\smartdiagramset{%
  back arrow disabled=true,
  module minimum width=2cm,
  module minimum height=2cm,
  module x sep=3cm,
  text width=2cm,
  additions={
    additional item offset=0.5cm,
    additional item border color=red,
    additional arrow color=red,
    additional item width=2cm,
    additional item height=2cm,
    additional item text width=3cm
  }
}

\begin{figure}[h]
  \centering
  \begin{tikzpicture}[yscale=0.9, auto,
      block/.style = {
        rectangle, draw=black, thick, text width=8em, text centered,
        rounded corners, minimum height=3em },
      block-sharp/.style = {
          rectangle, draw, thick, text width=8em, text centered, minimum
          height=3em },
      line/.style = { draw, thick, ->, >=stealth },
      line-dashed/.style = { draw, thick, ->, densely dotted, >=stealth }]
    \node [block-sharp] (llvmprogram) at (0, 0) {\textsf{Program in LLVM}};
	\node [rectangle split, draw, rectangle split parts=2, text
    width=8em,rounded corners, text centered, fill=blue!30] (instr) at (0, -2.75)
    {\textsf{\textbf{Instrumentation:}} \nodepart{second}
    \parbox{3cm}{\textsf{1. phase \\ 2. phase \\ \vdots}}};
    \node [block-sharp] (instrprogram) at (0, -5.5) {\textsf{Instrumented program in LLVM}};
    \node [block-sharp] (rules) at (-4.5, -2) {\textsf{Instrumentation\\ rules in JSON}};
    \node [block-sharp] (defs) at (-4.5, -4) {\textsf{Definitions of
    instrumentation functions in LLVM}};
    \node [block-sharp] (plugins) at (4.5, -2.75) {\textsf{Plugins}};
    % connect all nodes defined above
    \draw[line] (llvmprogram) -> (instr);
    \draw[line] (instr) -> (instrprogram);
    \draw[line] (rules) -> (instr);
    \draw[line] (defs) -> (instr);
    \draw[line] ($(instr.east)+(0, 0.15)$) -> ($(plugins.west)+(0, 0.15)$);
    \draw[line] ($(plugins.west)-(0, 0.15)$) -> ($(instr.east)-(0, 0.15)$);
    \draw[line] (llvmprogram) -| (plugins);
  \end{tikzpicture}
  \caption{The scheme of configurable instrumentation.}
  \label{fig:scheme}
\end{figure}
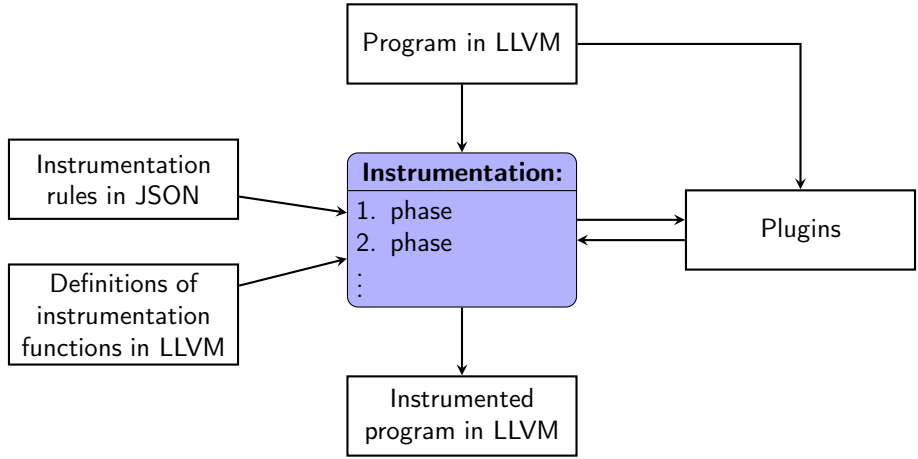

The basic schema of \instr tool is depicted in Figure~\ref{fig:scheme}.
% Since it works on top of LLVM, the given program that is supposed to
% be instrumented has to be translated to LLVM first.
Besides the LLVM bitcode to be instrumented, the tool needs to be
supplied with two files created by a user: a~file with definitions of
so called \emph{instrumentation functions} whose calls will be
inserted into the code, and a JSON file with \emph{instrumentation
  rules} that define how the LLVM bitcode should be instrumented with
calls of instrumentation functions. In practice, the fact that our
tool can insert just calls to instrumentation functions is not a
restriction as these functions can contain arbitrary code.
% We only allow to insert call
% instructions since it is sufficient as any other demanded instruction
% can be wrapped in a~called function.

The instrumentation proceeds in one or more phases, each phase defined by a set
of instrumentation rules. In each phase, the tool goes through all instructions of 
% the module loaded from
the given LLVM bitcode and it looks for instructions matching any
instrumentation rule of the current phase. If a match is found,
conditions of the instrumentation rule are evaluated. This is where
plugins can be queried. If conditions are satisfied, the rule is
applied, i.e., a~new code is inserted according to the rule and some
information can be gathered for a later use. The result of the
instrumentation process is again an LLVM bitcode.

% The phased instrumentation is useful because it allows the user to gather some
% information in earlier phases and use it in later phases. For example, it is
% pointless to insert a check for memory leaks if there is no call to
% \emph{malloc} in the program. Therefore, the presence of a call to
% \emph{malloc} can be noted in the first phase, and the check for memory leaks
% can be inserted in a later phase according to the noted information.

%------------------------------------------------------------------------------
\section{Instrumentation Rules}

Each instrumentation phase is defined by a set of instrumentation
rules.  An instrumentation rule consists of two parts saying
\emph{when} it should be applied and \emph{what} its effect is. The
first part of an instrumentation rule is specified by
\begin{itemize}
\item functions in which the rule is applied (typically \texttt{main}
  or all functions),
\item a sequence of instructions that should be matched, and
\item conditions under which the rule is applied.
\end{itemize}
The second part describes 
\begin{itemize}
\item the instrumentation function call that should be inserted,
\item where it should be inserted (before or after the matched sequence), and
\item information-gathering effects of the rule, namely setting flags
  and remembering values in an auxiliary list.
\end{itemize}
Moreover, there are two other kinds of rules, namely rules for
instrumentation of global variable declarations and rules
instrumenting entry and exit points of functions.

In the rest  of this section, we focus on  conditions. The other parts
of instrumentation rules are described in the following section.

\subsection{Conditions}\label{sec:conditions}

An instrumentation rule may contain conditions. To apply a rule, all
conditions of the rule have to be satisfied. 

Each condition consists of a \emph{query} and a list of \emph{expected
  results}. There are two kinds of queries, namely \emph{flag queries}
and \emph{plugin queries}.

Flags are typically used to pass some information between
instrumentation phases. They are set during application of instruction
rules. Subsequently, a rule can be conditioned by a certain value of a
specified flag. A flag query is simply the name of the flag. The query
is satisfied if the current value of the flag is in the list of
expected results.  For example, memory safety instrumentation sets the
flag \texttt{mallocPresent} to \emph{``true''} when instrumenting
a~call to \emph{malloc} in the first phase of the instrumentation. In
the second phase, it inserts a~check for memory leaks at the end of
\emph{main} only if the flag is set to \emph{``true''}, i.e.~if there
is a call to \emph{malloc} somewhere in the program.

Plugin queries are intended as an interface to static analyses.  In
general, a plugin query is formed by a keyword and some
parameters. These parameters are values taken from the program being
instrumented (typically the whole instruction matched by the rule or
just an operand of this instruction). A query is satisfied if some
plugin supporting this kind of query returns a string from the list
of expected results.  For example, a plugin query
\texttt{canOverflow(op)}, where \texttt{op} is an instruction for an
arithmetic operation with its operands, corresponds to the question
``Can the result of \texttt{op} overflow?''. Such a query can be
answered by a plugin performing a range analysis. If the analysis says
that the operation cannot overflow, the plugin answers
\emph{``false''} and it answers \emph{``maybe''} otherwise. Besides
the query \texttt{canOverflow(op)}, the range analysis plugin can also
evaluate the query \texttt{canBeZero(value)} corresponding to the
question whether \texttt{value} can be zero. Note that \texttt{value}
is a term of the LLVM terminology and it can refer, for example, to a
variable. We also support some queries evaluated by a pointer analysis
including the following ones:
\begin{description}
% \item[\texttt{hasKnownSize(pointer)}] Can we statically determine the
%   size of the memory object referenced by \texttt{pointer}?
\item[\texttt{isNull(pointer)}] Does the given \texttt{pointer} point to NULL?
\item[\texttt{isRemembered(value)}] Was the given value or a pointer
  to the value remembered in the auxiliary list by some previously
  applied instrumentation rule? This information needs to be evaluated
  by a pointer analysis plugin because we ask not only about the value
  itself, but also about pointers to the value. We use this query in
  the configuration for checking memory safety to determine whether a
  check of a specific pointer was inserted in an earlier phase of the
  instrumentation.
\item[\texttt{isValidPointer(addr, len)}] Is a dereference of \texttt{len}
  bytes starting from the address \texttt{addr} a valid operation? 
\end{description}
Plugins supporting these queries usually answer \emph{``true''},
\emph{``false''}, or \emph{``maybe''}. If a plugin does not implement some
query, it is not asked to evaluate it. The set of queries is not fixed and each
plugin can implement its own queries (without the need of changing the
\instr source code).

%------------------------------------------------------------------------------
\section{Configuration}\label{sec:config}

\lstset{
%    basicstyle=\footnotesize,
    string=[s]{"}{"},
    stringstyle=\color{blue},
    comment=[l]{:},
    commentstyle=\color{black},
    escapeinside={<@}{@>}
}

\begin{figure}[t]
\begin{lstlisting}
{
  "analyses": ["libRangeAnalysis.so"],
  "phases": [
     {
        "instructionsRules": [
           {
             "in": "*",
             "findInstructions": [
                {
                  "returnValue": "*",
                  "instruction": "sdiv",
                  "operands": ["*","<t1>"]
                }
             ],
             "conditions": [
                {
                  "query": ["canBeZero","<t1>"],
                  "expectedResults": ["true"]
                }
             ],
             "newInstruction": {
                  "instruction": "call",
                  "operands": ["<t1>","checkDivisionByZero"]
             },
             "where": "before"
           }
        ]
     }
  ]
}
\end{lstlisting}
\caption{Example of a JSON configuration file for inserting
  division-by-zero checks.}
\label{fig:example_config}
\end{figure}

In this section, we describe the basic structure of a JSON file
defining instrumentation rules. We illustrate the structure with the
configuration for inserting division-by-zero checks presented in
Figure~\ref{fig:example_config}. 

A JSON file with instrumentation rules may contain the following
fields, where only the field \texttt{phases} is mandatory.
\begin{center}
\begin{tabular}[h]{>{\bfseries}p{4.3cm} | p{9.2cm}}
%  \texttt{file}                & Path to the file with definitions of instrumentation functions. \\
%  \hline
  \texttt{analyses}            & List of paths to plugins. \\
  \hline
  \texttt{flags}               & List of flags that can be set during instrumentation. \\
  \hline \texttt{phases}       & List of instrumentation phases. Each
                                 phase contains a list of \texttt{instructionsRules}
                                 and/or \texttt{globalVariablesRules}. 
                                 The phases are processed
                                 in the order given by their position in the list. \\
  % \hline
  % \texttt{globalVariablesRules} & List of rules for instrumenting global
  %                                variables.
\end{tabular}
\end{center}

Instrumentation rules may use \emph{configuration
  variables}. Syntactically, these variables are enclosed by the
\texttt{<} and \texttt{>} characters. Configuration variables serve to
store some parts of matched instructions (e.g., their operands) for
use in conditions, inserted function calls, and to be stored in the
auxiliary list.

% For example, to store an operand of a \texttt{load} instruction to
% the variable \texttt{<t>}, a field \texttt{operands} that will
% correspond to the operands of \texttt{load} instruction will be set
% to \texttt{["<t>"]}.  Variable \texttt{<t>} will contain the value
% of the operand of \texttt{load} instruction in the scope of the
% application of the current rule and it can be used e.g.~as an
% argument of the function call that will be inserted or as an
% argument of a condition.

\subsection{\texttt{instructionsRules}}

Each element of \texttt{instructionsRules} is a JSON object described by 
several fields:
\begin{description}
\item[\texttt{in}] Name of a function, in which this rule should be
  considered for application. For example, \texttt{"in": main} means
  that this rule will be applied only in the \emph{main} function. It
  can also be set to "*" meaning that it should be taken into
  consideration in all functions (see
  Figure~\ref{fig:example_config}).

\item[\texttt{findInstructions}] Sequence of instructions we are
  searching for. For each instruction in the sequence, we need to fill
  in a field \texttt{instruction} that specifies the name of the
  instruction to be matched. The field \texttt{returnValue} allows to
  remember the return value of the instruction in a given
  configuration variable. It can be omitted or set to "*" if the
  return value is not needed. Finally, the field \texttt{operands}
  enables either to match the operands or to remember them in
  configuration variables. We can also optionally fill in the field
  \texttt{getTypeSize}.  This field can be used only with
  \texttt{load}, \texttt{store} or \texttt{alloca} instructions and it
  stores the size of the type of the value that is being loaded,
  stored, or allocated to the given configuration variable.

  An example of a \texttt{findInstructions} field can be found in
  Figure~\ref{fig:example_config}. We are searching for a sequence of
  instructions of length one, namely for a \texttt{sdiv} instruction.
  In this case, we do not care about the return value, so the
  \texttt{returnValue} field is set to "*". The second operand of
  \texttt{sdiv} will be stored in the configuration variable
  \texttt{<t1>} (the first operand will be skipped as it is set to
  "*").

\item[\texttt{conditions}] An optional field specifying a list of
  conditions that have to be satisfied in order to apply the rule (see
  Section~\ref{sec:conditions}). A condition consists of the fields
  \texttt{query} and \texttt{expectedResults}. The \texttt{query} is a
  list where the first element is the name of a query and other
  elements are parameters passed to the query. The
  \texttt{expectedResults} is a list of expected results of the query.

  In Figure~\ref{fig:example_config}, the \texttt{conditions} list
  contains one condition that has to be satisfied for the rule to be
  applied. It is a plugin query that will be satisfied if the plugin
  loaded from the shared object \texttt{libRangeAnalysis.so} will
  answer \emph{``true''} to the query \texttt{canBeZero} with the
  value of \texttt{<t1>} passed as a parameter. The following code
  shows a condition with a flag query that is satisfied if the flag
  \texttt{mallocPresent} is set to \emph{``true''}.

  \begin{minipage}{\linewidth}
    \lstset{
% basicstyle=\footnotesize,  
      string=[s]{"}{"},
      stringstyle=\color{blue}, comment=[l]{:},
      commentstyle=\color{black}, }
\begin{lstlisting}
"conditions": [
   {
      "query":["mallocPresent"],
      "expectedResult":["true"]
   }
]
\end{lstlisting}
  \end{minipage}

\item[\texttt{newInstruction}] A new instruction that is to be
  inserted. It contains two mandatory fields: \texttt{instruction}
  that specifies the name of the new instruction (for now, only
  \texttt{call} instructions are supported), and \texttt{operands} of
  the instruction. The last operand is the name of the function that
  should be called. For example, the instrumentation rule described in
  Figure~\ref{fig:example_config} will insert a
  call to the function \texttt{checkDivisionByZero} with the argument 
  stored in \texttt{<t1>}.
 
\item[\texttt{where}] The location of insertion. It can be
  \emph{``before''} or \emph{``after''} the found sequence of
  instructions. Alternatively, it can have the value \emph{``entry''}
  or \emph{``return''} saying that the code should be inserted at the
  entry point or before every \texttt{return} instruction of the
  current function, respectively. If the value is \emph{``entry''} or
  \emph{``return''}, the rule does not need to contain the
  \texttt{findInstructions} field (as it is ignored).

\item[\texttt{setFlags}] This optional field describes the list of
  pairs \texttt{[flag, string]} that sets each \texttt{flag} to the
  corresponding \texttt{string} if the rule was applied. The following
  code sets the flag \texttt{loadFlag} to \emph{``true''} and the flag
  \texttt{testFlag} to \emph{``false''}.

  \begin{minipage}{\linewidth}
    \lstset{
%            basicstyle=\footnotesize,
            string=[s]{"}{"},
            stringstyle=\color{blue},
            comment=[l]{:},
            commentstyle=\color{black}
    }
\begin{lstlisting}
"setFlags":[["loadFlag","true"],["testFlag","false"]]
\end{lstlisting}
  \end{minipage}

\item[\texttt{remember}] An optional field specifying the name of a
  configuration variable whose value will be stored in the auxiliary
  list if the rule is applied.
 % If some rule is
 %  conditioned by a~query \texttt{isRemembered} with the variable or a
 %  pointer to the variable as its parameter, the condition will be
 %  satisfied.
\end{description}

\subsection{\texttt{globalVariablesRules}}

\texttt{globalVariablesRules} can be used to instrument declarations
of global variables. The original motivation was in memory safety
checking, where we needed to insert functions tracking memory blocks
corresponding to global variables.
Each element of \texttt{globalVariablesRules} is a JSON
object with the following fields:
\begin{description}
\item[\texttt{findGlobals}] Contains a mandatory field
  \texttt{globalVariable} that stores the address of the global
  variable to the given configuration variable, and an optional field
  \texttt{getTypeSize} that gets the size of the type of the global
  variable. In the following example, the field \texttt{findGlobals}
  stores the address of a global variable in \texttt{<t1>} and stores
  the size of its type in \texttt{<t2>}.

  \begin{minipage}{\linewidth}
        \lstset{
%            basicstyle=\footnotesize,
            string=[s]{"}{"},
            stringstyle=\color{blue},
            comment=[l]{:},
            commentstyle=\color{black},
        }
\begin{lstlisting}
"findGlobals": {
   "globalVariable": "<t1>",
   "getTypeSize": "<t2>"
}
\end{lstlisting}
  \end{minipage}
\item[\texttt{conditions}] The same as in \texttt{instructionsRule}.
\item[\texttt{newInstruction}] The same as in
  \texttt{instructionsRule}.
\item[\texttt{in}] The name of the function, at the beginning of which
  the new instruction will be inserted. This time, the field cannot be
  set to "*".
\end{description}

\section{Implementation and Usage}
The \instr tool can be run by executing the compiled binary \texttt{sbt-instr}:

\begin{center}
\texttt{./sbt-instr <config> <input> <definitions> <output>}
\end{center}
where the arguments have the following meaning:

\begin{itemize}
\item \texttt{<config>} is a file with instrumentation rules in JSON as described in Section~\ref{sec:config},
\item \texttt{<input>} is the LLVM bitcode to be instrumented,
\item \texttt{<definitions>} is the LLVM bitcode with definitions of instrumentation functions, and
\item \texttt{<output>} is the name of the output file.
\end{itemize}

As we primarily use the instrumentation for programs written in the C
programming language, there is a script \texttt{sbt-instr-c} that works
directly with C programs:

\begin{center}
\texttt{./sbt-instr-c <config> <input> [--output=<output>]}
\end{center}

The arguments have the same meaning as for \texttt{sbt-instr}, but now the
\texttt{<input>} is a program in LLVM or C. The script compiles the
program automatically to LLVM if needed and the final instrumented LLVM bitcode
is stored to \texttt{<output>} file (if given) or into \texttt{out.bc}.
Note that there is no argument for definitions of instrumentation functions.
The path to the file with definitions is specified by
a new top-level field \texttt{file} in the JSON configuration.
The definitions of instrumentation functions can be given in LLVM or
in C (in which case they are automatically compiled into LLVM).

The \instr tool is distributed as a part of the toolbox called
\symbiotic~\cite{symbiotic5} and comes with two predefined
configurations used for program analysis, namely a configuration for checking
memory safety~\cite{spin2018} briefly described in Section~\ref{sect:introduction}
and a configuration for checking signed integer
overflows. The latter configuration inserts a check before every binary
operation over signed integers
that may potentially overflow based on the results of a range analysis.

%\textcolor{red}{null-check config? division-by-zero config?}

\instr tool together with the predefined configurations for checking memory
safety and integer overflows can be found at
\begin{center}
{
\color{blue}
\url{https://github.com/staticafi/sbt-instrumentation}
}
\end{center}
under Apache 2.0 license. The tool is implemented in the C++ programming language.
It uses an open-source parser for the JSON format
JsonCpp\footnote{\url{https://github.com/open-source-parsers/jsoncpp}}. The
\instr repository also contains the script \texttt{sbt-instr-c} written in
Python.

To compile and run the tool, it is necessary to have CMake (minimal version
2.8.8) and the LLVM (minimal version 3.9.1) together with Clang installed. To
use the predefined configurations for checking memory safety or integer overflows,
it is also needed to install the
\texttt{dg} library\footnote{\url{https://github.com/mchalupa/dg}}
for pointer analysis or the \texttt{ra} library\footnote{\url{https://github.com/xvitovs1/ra}} for range analysis, respectively.

%------------------------------------------------------------------------------
\section{Related Work}

Even though most of the tools for a program analysis implement their own
instrumentation aimed directly at their purposes, there are a few papers about
general-purpose instrumentation. However, none of these tools supports
instrumentation of LLVM bitcode and they do not provide the full functionality
of our tool, in particular, the interface to static analyses.

\textsc{ATOM}~\cite{atom} is a tool for a binary instrumentation that allows a
user to describe how the code will be instrumented and to define the code that
will be inserted. The user has to implement their own instrumentation routine
with the help of ATOM's C-language user interface. This routine defines the way
of iterating over instructions and when to insert new code.
\textsc{FIT}~\cite{fit} is a binary instrumentation tool that is
backwards compatible with \textsc{ATOM}, but it supports more different architectures.

Geimer et al.~\cite{GeimerSMW09} propose an approach for general entry/exit
instrumentation and explore the basic constructs that are needed for
a configurable instrumentation routine. They evaluate the proposed
approach by implementing a prototype based on the instrumentation tool from the
\textsc{TAU} performance-analysis framework~\cite{tau2006} which currently
works with C, C++, and Fortran programs.

Stephenson et al.~\cite{StephensonHLEJN15} designed a flexible instrumentation
framework called \textsc{SASSI} for NVIDIA GPUs. The motivation was to allow
injecting user-defined code into GPU programs instead of choosing from
predefined profiling functions. The user therefore can specify \textit{what}
will be inserted and \textit{where} it will be inserted.

%------------------------------------------------------------------------------
\section{Conclusion}

We have introduced \instr, a tool for general-purpose configurable
instrumentation of LLVM bitcode. The instrumentation process can be divided
into multiple phases. Every phase can gather information about what has been already
instrumented and this information can be used by subsequent phases.
Furthermore, the tool provides an extensible API for static analyses
in the form of plugins. Injecting new code can be then conditioned by
the results of external static analyses.

We described the workflow of the tool and the format for
configuration of the instrumentation process.
\instr was integrated into the \symbiotic toolbox
where it plays a crucial role for memory safety checking.

In the future, we plan to develop more configurations and try to plug in
new static analyses. We also want to add support for replacing parts of
the matched code.

%In the future, we plan to make a few adjustments to \instr tool. For example,
%we are thinking about enabling not only insertion of instruction but also their
%replacement. We would also like to make some changes to the condition system in
%the JSON configuration file. Currently, the conditions have to be evaluated by
%all the plugins named in the configuration, which can be inconvenient in some
%cases. Therefore, we plan to add a possibility to specify the plugin that will
%be queried. Another idea we are considering is to completely change the form of
%the condition system to a decision tree.

%------------------------------------------------------------------------------

\label{sect:bib}
\bibliographystyle{unsrt}
\bibliography{lpar}

\begin{thebibliography}{10}

\bibitem{Serebryany2012}
Konstantin Serebryany, Derek Bruening, Alexander Potapenko, and Dmitry Vyukov.
\newblock {AddressSanitizer: A Fast Address Sanity Checker}.
\newblock In {\em Proceedings of the 2012 USENIX Conference on Annual Technical
  Conference}, USENIX ATC'12, pages 28--28. USENIX Association, 2012.

\bibitem{JProfiler}
{Java Profiler - JProfiler}.
\newblock
  \url{https://www.ej-technologies.com/products/jprofiler/overview.html}.
\newblock Accessed: 2018-07-24.

\bibitem{cbmc}
Daniel Kroening and Michael Tautschnig.
\newblock {CBMC} - {C} bounded model checker - (competition contribution).
\newblock In {\em Tools and Algorithms for the Construction and Analysis of
  Systems - 20th International Conference, {TACAS} 2014, Held as Part of the
  European Joint Conferences on Theory and Practice of Software, {ETAPS} 2014,
  Grenoble, France, April 5-13, 2014. Proceedings}, volume 8413 of {\em Lecture
  Notes in Computer Science}, pages 389--391. Springer, 2014.

\bibitem{thesis}
Martina Vitovsk{\'{a}}.
\newblock {Instrumentation of LLVM IR}.
\newblock Master's thesis, Masaryk University, Faculty of Informatics, Brno,
  2018.

\bibitem{llvm}
The {LLVM} compiler infrastructure.
\newblock \url{http://llvm.org}, 2017.

\bibitem{spin2018}
Marek Chalupa, Jan Strej\v{c}ek, and Martina Vitovsk{\'{a}}.
\newblock Joint forces for memory safety checking.
\newblock In {\em Model Checking Software - 25th International Symposium,
  {SPIN} 2018, Malaga, Spain, June 20-22, 2018, Proceedings}, volume 10869 of
  {\em Lecture Notes in Computer Science}, pages 115--132. Springer, 2018.

\bibitem{symbiotic5}
Marek Chalupa, Martina Vitovsk{\'{a}}, and Jan Strej\v{c}ek.
\newblock Symbiotic 5: Boosted instrumentation (competition contribution).
\newblock In {\em Tools and Algorithms for the Construction and Analysis of
  Systems - 24th International Conference, {TACAS} 2018, Held as Part of the
  European Joint Conferences on Theory and Practice of Software, {ETAPS} 2018,
  Thessaloniki, Greece, April 14-20, 2018, Proceedings, Part {II}}, volume
  10806 of {\em Lecture Notes in Computer Science}, pages 442--446. Springer,
  2018.

\bibitem{atom}
Amitabh Srivastava and Alan Eustace.
\newblock {ATOM} - {A} system for building customized program analysis tools.
\newblock In {\em Proceedings of the {ACM} SIGPLAN'94 Conference on Programming
  Language Design and Implementation (PLDI), Orlando, Florida, USA, June 20-24,
  1994}, pages 196--205. {ACM}, 1994.

\bibitem{fit}
Bruno~De Bus, Dominique Chanet, Bjorn~De Sutter, Ludo~Van Put, and Koen~De
  Bosschere.
\newblock The design and implementation of {FIT:} a flexible instrumentation
  toolkit.
\newblock In {\em Proceedings of the 2004 {ACM} {SIGPLAN-SIGSOFT} Workshop on
  Program Analysis For Software Tools and Engineering, PASTE'04, Washington,
  DC, USA, June 7-8, 2004}, pages 29--34. {ACM}, 2004.

\bibitem{GeimerSMW09}
Markus Geimer, Sameer Shende, Allen~D. Malony, and Felix Wolf.
\newblock A generic and configurable source-code instrumentation component.
\newblock In {\em Computational Science - {ICCS} 2009, 9th International
  Conference, Baton Rouge, LA, USA, May 25-27, 2009, Proceedings, Part {II}},
  volume 5545 of {\em Lecture Notes in Computer Science}, pages 696--705.
  Springer, 2009.

\bibitem{tau2006}
Sameer Shende and Allen~D. Malony.
\newblock {The Tau Parallel Performance System}.
\newblock {\em {IJHPCA}}, 20(2):287--311, 2006.

\bibitem{StephensonHLEJN15}
Mark Stephenson, Siva Kumar~Sastry Hari, Yunsup Lee, Eiman Ebrahimi, Daniel~R.
  Johnson, David~W. Nellans, Mike O'Connor, and Stephen~W. Keckler.
\newblock Flexible software profiling of {GPU} architectures.
\newblock In {\em Proceedings of the 42nd Annual International Symposium on
  Computer Architecture, Portland, OR, USA, June 13-17, 2015}, pages 185--197.
  {ACM}, 2015.

\end{thebibliography}
%------------------------------------------------------------------------------
\appendix
%------------------------------------------------------------------------------
% Index
%\printindex

%------------------------------------------------------------------------------
\end{document}